\documentclass[fleqn,twoside]{article}
\usepackage{espcrc2}
\usepackage{graphicx}

\title{A Feynman diagram analyzer DIANA: recent development
       \thanks{Supported by INTAS-CERN 99-0377}}

\author{M.~Tentyukov,\address{BLTP JINR, Dubna,
        Russia}\thanks{Supported by DFG under the project FL 241/4-2}
        J.~Fleischer\address{Universit\"at Bielefeld,
        Fakult\"at f\"ur Physik, Bielefeld, Germany},%
        }
\begin{document}

\begin{abstract}
New developments concerning the extension of the Feynman diagram analyzer
DIANA are presented. We discuss new graphic facilities, application of DIANA
to processes with Majorana fermions and different approaches to automation
of momenta distribution.
\vspace{1pc}
\end{abstract}

\maketitle

The project called DIANA (DIagram ANAlyzer)\cite{Diana} for the evaluation
of Feynman diagrams was started by our group some time ago. It was already
used to calculate several processes \cite{usage}. The recent
development\footnote[1]{For details look at\\
http://www.physik.uni-bielefeld.de/$\tilde{\,}$tentukov/diana.html}
of this project will be shortly described below.

The pictorial representation of diagrams described in \cite{graf}
includes three different kinds of postscript files. Now one more
kind is available, the encapsulated postscript file containing
particle lines together with momenta flows. Particle identifiers now can be
depicted by different fonts, sizes and colours:

\begin{center}
\vspace*{-0.2cm}
\includegraphics[width=4.0cm]{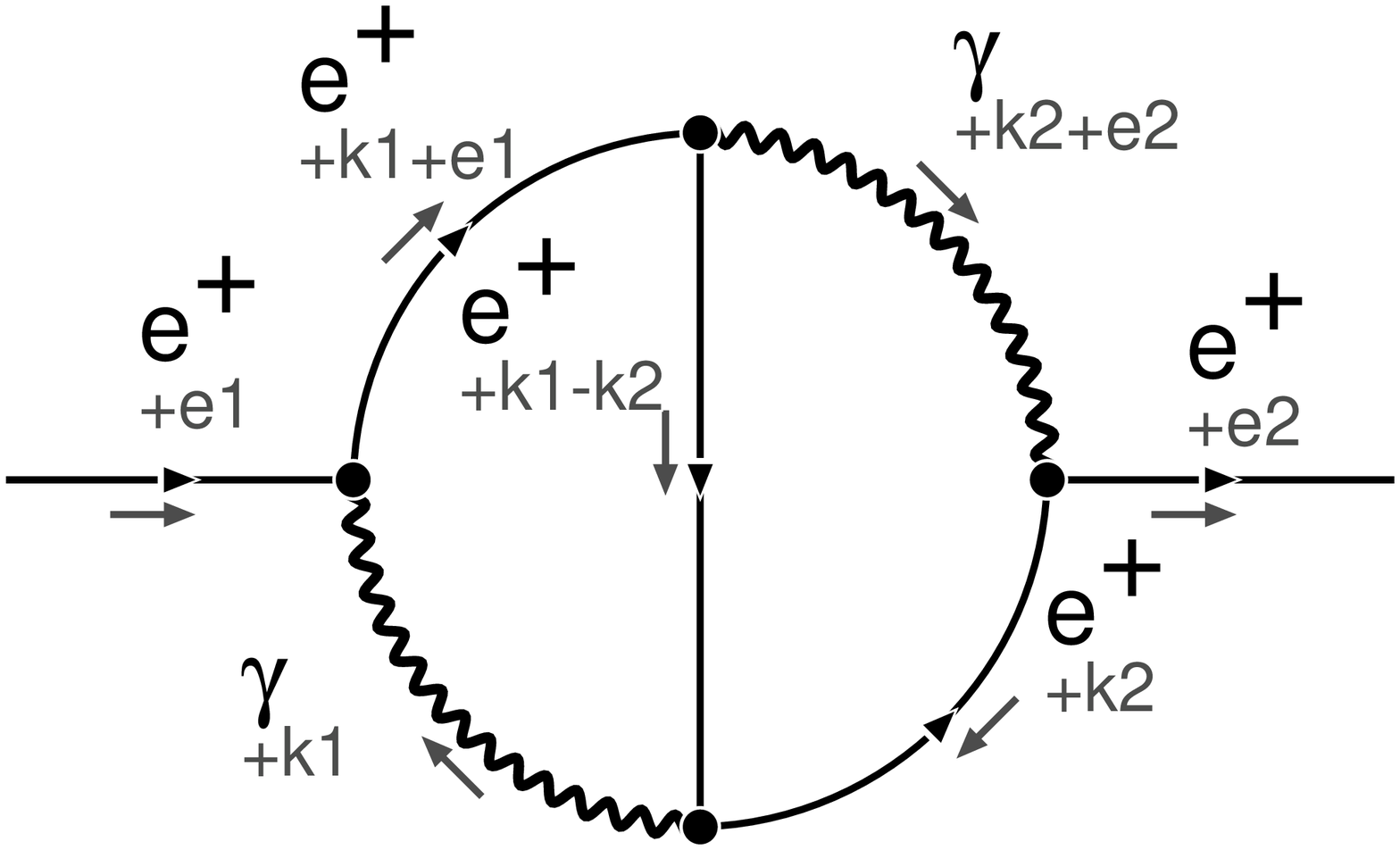}
\end{center}
\vskip -0.2cm

DIANA has several features
to support simple algorithmic Feynman rules for interactions with Majorana
fields, elaborated by Denner {\em et al} \cite{Majorana}.
The basic idea is the following: a
new keyword in the model description of propagators and vertices is introduced.
Its
default value \verb|fflow| may be changed by
the directive \verb|fermion number flow = newkw|.
This identifier is supposed to take the values 0, -1 or 1. $\pm 1$ stand for
Dirac fermions and 0 for all other particles. For vertices it will be
-1, if at least one incident line has -1. The assignment of these values is
performed by DIANA.

For example, the Lagrangian $\bar\lambda\Gamma\psi W^- +
\bar\psi\bar\Gamma\lambda W^+$, ($\psi$=\verb|f| -- Dirac fermion,
$\lambda$=\verb|l| --
Majorana fermion, $W^\pm$=(\verb|Wp,Wm|) -- charged vector, $\Gamma$ is a Gamma matrix)
in terms of a DIANA model reads in terms of propagators
\begin{verbatim}
[Wp,Wm;a;ww(num,ff=fflow,fn=fnum);0]
[f,F  ;f;ff(num,ff=fflow,fn=fnum);0]
[l,l  ;l;ll(num,ff=fflow,fn=fnum);0]
\end{verbatim}
and vertices
\begin{verbatim}
[l,f,Wm;; lfWm(num,ff=fflow,fn=fnum)]
[F,l,Wp;; FlWp(num,ff=fflow,fn=fnum)].
\end{verbatim}

Process $(\psi,\bar\psi) \to (\psi,\bar\psi)$, diagram ``number 3'':

\begin{center}
\includegraphics[width=4.5cm]{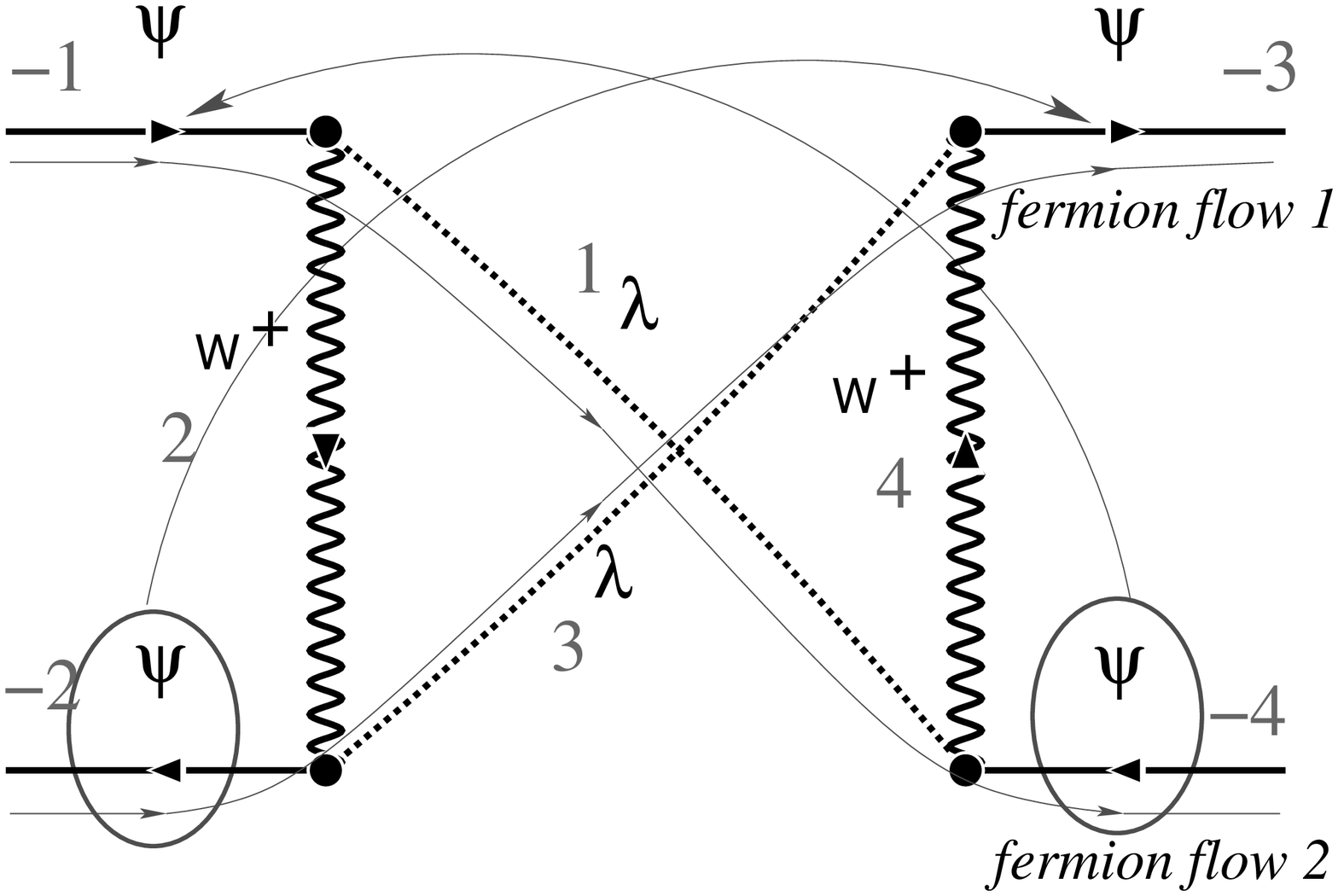}
\end{center}

Generated FORM input:
\begin{verbatim}
G Rq =
  spl(-3,ff=1,fn=1)*spl(-4,ff=-1,fn=2)*
  (-1)*FlWp(3,ff=1,fn=1)*ll(3,ff=0,fn=1)*
  FlWp(2,ff=-1,fn=1)*lfWm(4,ff=-1,fn=2)*
  ll(2,ff=0,fn=2)*lfWm(1,ff=1,fn=2)*
  ww(1,ff=0,fn=0)*ww(4,ff=0,fn=0)
  *spr(-2,ff=-1,fn=1)*spr(-1,ff=1,fn=2);
\end{verbatim}

Topologies \cite{Diana} are represented in terms of
    ordered pairs of numbers like \verb|(fromvertex, tovertex)|.
    All external legs have negative numbers.

Sometimes the number of topologies is too large such that
it is impossible to assign
momenta to the lines in all topologies ``by hand''.

Of course, momenta can be introduced automatically. The user may specify loop
momenta via the macro \verb|\loopmomenta| in the ``create'' file, e.g.
\verb|\loopmomenta(k1,k2,k3)|,
and DIANA will assign momenta automatically using ``$k1$'', ``$k2$'' and
``$k3$'' as the
loop integration momenta.

Sometimes it is important to keep some definite lines free from the
external momenta.
If the users specifies \verb|SET _MARK_LOOP=YES| in the
``create'' file, the topology editor will be invoked in a special mode, and
the user can point out which lines should carry bare
integration momenta. All remaining momenta will be assigned automatically.

Sometimes it is
necessary to use more sophisticated distribution.
For example, the user may
want to use his favorite momenta like $k-p_1, k-p_2, k-p_3$, etc. assigned to
some definite lines (see e.g. \cite{FTJ9907327}). For such cases, DIANA
provides a possibility to define momenta only for the virtual lines, and
the full set of topologies will be defined from these ``internal'' topologies
attaching the external legs. Momenta for
internal parts are defined in terms of combinations of loop momenta ($k$ in
the above example) and some ``abstract'' tokens ($p_1, p_2$ etc.), and for
each topology DIANA may express these tokens in terms of external momenta.

Another example: often topologies are generated by more complicated
(``generic'') ones by scratching lines. In such cases one wants to stick to
the momenta introduced for the lines which are kept. E.g. the user
investigates the generic topology
\begin{verbatim}
generictopology A =
 (-2,2)(-1,1)(1,3)(3,2)(2,4)(4,1)(3,4):
 p1+k2,p1+k1,k1,k2,k2-k1.
\end{verbatim}
Then DIANA will generate topologies form this one
by scratching lines in the following manner:
\begin{center}
\includegraphics[width=5.5cm]{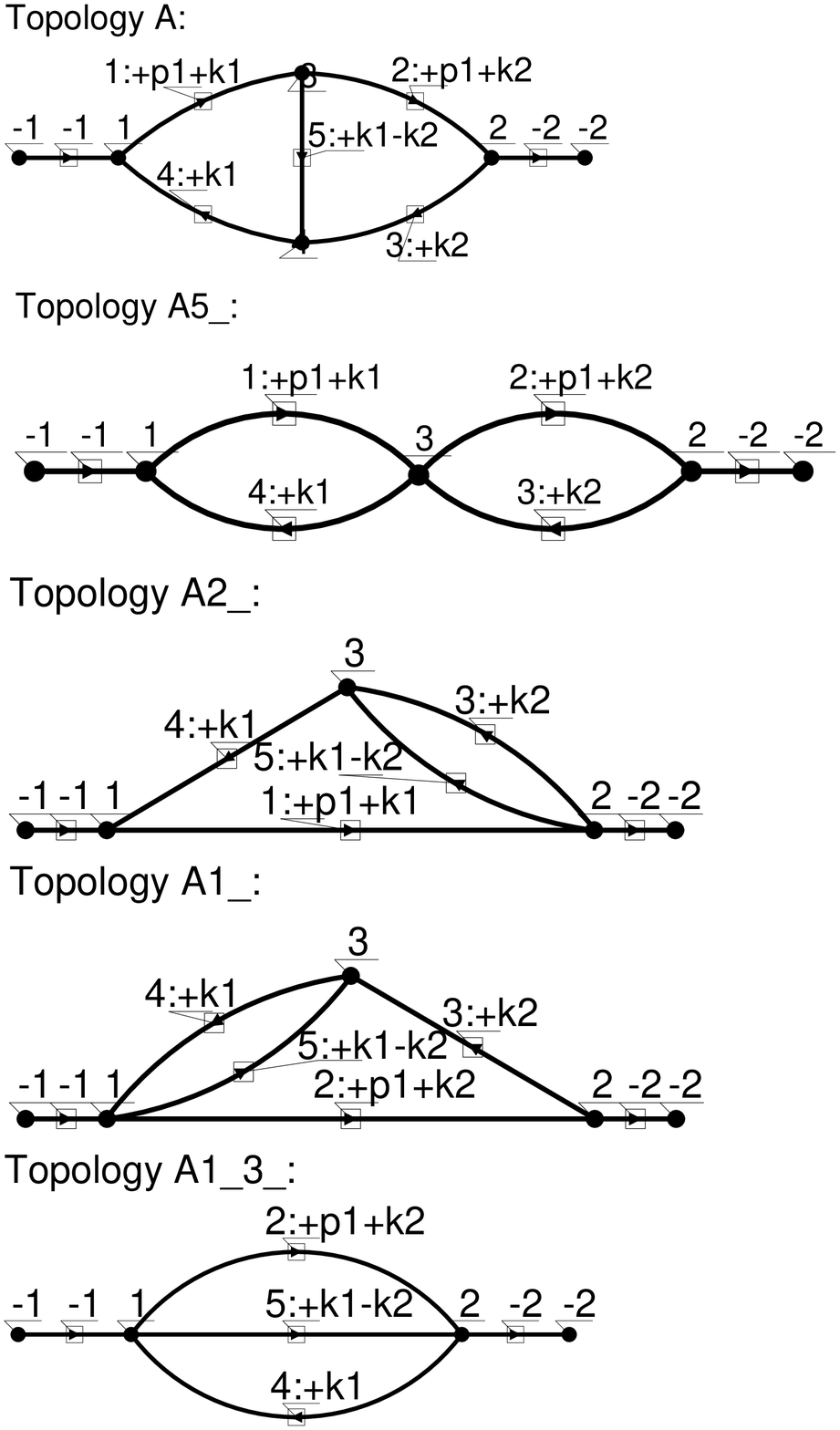}
\end{center}

\end{document}